\documentclass[10pt, a4paper, twocolumn]{IEEEtran}
\usepackage{amsmath,epsfig,amssymb,verbatim,amsopn,cite,subfigure,multirow}
\usepackage{balance,nopageno}
\usepackage{algorithm,algorithmic}
\usepackage[usenames,dvipsnames]{color}
\usepackage[all]{xy}  %%% used to make block diagram
\usepackage{url}
\usepackage{amsfonts}
\usepackage{amssymb}
\usepackage{epsfig}
\usepackage{epstopdf}
\usepackage{slashbox}
\usepackage{bm}
\usepackage{balance}
\usepackage[margin=17.0mm,top=17mm,bottom=17mm]{geometry}%margins; tweak for printer deficiencies
\newtheorem{Remark}{Remark}

\newtheorem{proposition}{Proposition}

\newcommand{\E}[1]{\mathbb{E}\left\{#1\right\}}

\newcommand{\qg}{{\bf g}}
\newcommand{\qh}{{\bf h}}

\newcommand{\qn}{{\bf n}}

\newcommand{\qp}{{\bf p}}

\newcommand{\qr}{{\bf r}}
\newcommand{\qs}{{\bf s}}

\newcommand{\qw}{{\bf w}}
\newcommand{\qx}{{\bf x}}
\newcommand{\qy}{{\bf y}}

\newcommand{\qA}{{\bf A}}
\newcommand{\qB}{{\bf B}}
\newcommand{\qC}{{\bf C}}
\newcommand{\qD}{{\bf D}}

\newcommand{\qG}{{\bf G}}
\newcommand{\qH}{{\bf H}}
\newcommand{\qI}{{\bf I}}

\newcommand{\qN}{{\bf N}}

\newcommand{\qQ}{{\bf Q}}

\newcommand{\qS}{{\bf S}}

\newcommand{\qW}{{\bf W}}
\newcommand{\qX}{{\bf X}}
\newcommand{\qY}{{\bf Y}}

\newcommand{\SI}{{\mathsf{SI}}}

\newcommand{\Eche}{\boldsymbol{\mathcal{{E}}}}

\newcommand{\Gad}{\qG_{\AP\dl}}

\newcommand{\Dad}{\qD_{\AP\dl}}
\newcommand{\Nad}{\qN_{\AP\dl}}
\newcommand{\Ead}{\Eche_{\AP\dl}}

\newcommand{\Gadhat}{\hat{\qG}_{\AP\dl}}
\newcommand{\Gdahat}{\hat{\qG}_{\dl\AP}}

\newcommand{\HLI}{{\qH}_{\SI}}
\newcommand{\HLIhat}{\hat{\qH}_{\SI}}
\newcommand{\ELI}{\Eche_{\SI}}

\newcommand{\MRC}{\mathsf{MRC}}
\newcommand{\MRT}{\mathsf{MRT}}

\newcommand{\Ds}{\mathsf{d}}
\newcommand{\AP}{\mathsf{a}}

\newcommand{\RdkIP}{{R}_{\mathsf{dl,k}}^{\mathsf{IP}}}
\newcommand{\RukhatIP}{\tilde{R}_{\mathsf{ul,k}}^{\mathsf{IP}}}

\newcommand{\Rdk}{{{R}}_{\mathsf{dl,k}}}
\newcommand{\Ruk}{{R}_{\mathsf{ul,k}}}

\newcommand{\Rdkhat}{\tilde{{R}}_{\mathsf{dl,k}}}
\newcommand{\Rukhat}{\tilde{{R}}_{\mathsf{ul,k}}}

\newcommand{\Hbarhat}{\hat{\bar{\qH}}}
\newcommand{\Hbar}{{\bar{\qH}}}

\newcommand{\Nrx}{N_{\mathsf{dl}}}
\newcommand{\Ntx}{N_{\mathsf{ul}}}
\newcommand{\Kul}{K_{\mathsf{ul}}}
\newcommand{\Kdl}{K_{\mathsf{dl}}}
\newcommand{\dl}{{\mathsf{d}}}
\newcommand{\ul}{{\mathsf{u}}}
\newcommand{\En}{{\mathsf{E}}}

\newcommand{\Plus}{\boldsymbol{\Phi}_{\dl}}
\newcommand{\PhE}{\boldsymbol{\Phi}_{\En}}

\newcommand{\Sn}{\sigma_n^2}
\newcommand{\Sap}{\sigma_{\SI}^2}
\newcommand{\Sadk}{\sigma^2_{\AP\dl,k}}

\newcommand{\diag}{\mathsf{diag}}

\newcommand{\be}{\begin{equation}} \newcommand{\ee}{\end{equation}}
\newcommand{\bea}{\begin{eqnarray}} \newcommand{\eea}{\end{eqnarray}}

\usepackage{multibib}
\usepackage[nodisplayskipstretch]{setspace}
\newcites{Prim}{Very important papers}

\definecolor{light-gray}{gray}{0.65}

\newcounter{mytempeqcounter}

\newcommand{\bigformulatop}[2]{%
  \begin{figure*}[!t]
    \normalsize
    \setcounter{mytempeqcounter}{\value{equation}}
    \setcounter{equation}{#1}
    #2

    \setcounter{equation}{\value{mytempeqcounter}}
    \hrulefill
    \vspace*{4pt}
  \end{figure*}
}

\title{\fontsize{0.84cm}{1cm}\selectfont Wireless Information and Power Transfer in Full-Duplex Systems
with Massive Antenna Arrays}

\author{{Mohammadali Mohammadi$^\dag$, Batu K. Chalise$^\ddag$, Himal A. Suraweera$^\S$, and Zhiguo Ding$^*$}\\  
\small{
$^\dag$Faculty of  Engineering, Shahrekord University, Iran (e-mail: m.a.mohammadi@eng.sku.ac.ir)\\
$^\ddag$Department of Electrical Engineering and Computer Science, Cleveland State University, USA (email: b.chalise@csuohio.edu)\\
$^\S$Department of Electrical and Electronic Engineering, University of Peradeniya, Sri Lanka (e-mail: himal@ee.pdn.ac.lk)\\
$^*$School of Computing and Communications, Lancaster University, UK (e-mail: z.ding@lancaster.ac.uk)
}}\normalsize

\begin{document}

\maketitle
\thispagestyle{empty}

\begin{abstract}
We consider a multiuser wireless system with a full-duplex  hybrid access point (HAP) that transmits to a set of users in the
downlink channel, while receiving data from a set of energy-constrained sensors in the uplink  channel. We assume that the HAP is equipped with a massive antenna array, while all users and sensor nodes have a single antenna. We adopt a time-switching protocol where in the first phase, sensors are powered through wireless energy transfer from HAP and HAP estimates the downlink channel of the users. In the second phase, sensors use the harvested energy to transmit to the HAP. The downlink-uplink sum-rate region is obtained by solving downlink sum-rate maximization problem under a constraint on uplink sum-rate. Moreover, assuming perfect and imperfect channel state information, we derive expressions for the achievable uplink and downlink rates in the large-antenna limit and approximate results that hold for any finite number of antennas. Based on these analytical results, we obtain the power-scaling law and  analyze the effect of the number of antennas on the cancellation of intra-user interference and the self-interference.
\end{abstract}
%***************************************************************************
%%***************************************************************************
\section{Introduction}
%***************************************************************************
%%***************************************************************************
Full-duplex (FD) wireless allows simultaneous transmission and reception of signals using the same frequency. Therefore, it has been touted as a promising technology to achieve increased spectral efficiency requirements of future 5G networks~\cite{Ashutosh:JSAC:2014}. In terms of practical FD implementation, the effect of self-interference (SI) due to the coupling of own high powered transmit signals to the receiver must be reduced. A variety of SI cancellation solutions have been reported in the recent literature to make FD implementation a near-term reality~\cite{Riihonen:JSP:2011,Sabharwal:TWC2012}.

On the other hand, popularity of multimedia centric wireless applications have created a high demand for energy. In contemporary networks, terminals are either connected to the electrical grid or rely on batteries for operation. Hence, limited operational life time of wireless networks has become an issue. As a potential solution, wireless nodes can be powered by harvesting ambient power or by wireless power transfer (WPT). The later approach especially becomes useful in sensor applications since a significant amount of energy can be harvested due to controllable fashion of WPT.

One potential application of FD is to use it at an access point (AP) for simultaneous uplink information reception and downlink energy delivery. In the context of a wireless-powered communication network (WPCN), such operation can be modified to consider a FD hybrid AP (HAP) with uplink information reception and downlink energy transfer. Several papers have investigated the information and energy transfer performance of such WET-enabled HAP systems with half-duplex (HD) operation~\cite{R.Zhang:JSAC:.2015,Long:TWC:2016} and FD operation ~\cite{HJu:TCOM:2014,Yongbo:2016,Nguyen:Eusipco2016}. In~\cite{R.Zhang:JSAC:.2015}, a throughput maximization problem for a WPT enabled massive multiple-input-multiple-output (MIMO) system consisting of a HAP and multiple single-antenna users has been investigated. In~\cite{Long:TWC:2016}, harvested energy maximization subject to rate requirements of information users has been considered for a massive MIMO WPCN. In~\cite{HJu:TCOM:2014}, resource allocation in a WPCN where a FD HAP is used for downlink energy broadcasting and uplink information reception has been studied. An optimal transmit power and time allocation problem for a single antenna FD WPCN has been solved in~\cite{Yongbo:2016}. A FD multisuer MIMO system has been studied in~\cite{Nguyen:Eusipco2016} where uplink users first harvest energy via BS energy beamforming before transmitting their information to the BS while at the same time the BS transmits information to the users in the downlink channel.

%without the effect of SI

In this paper, we consider a FD system in which a HAP performs WPT to a set of sensors while at the same time users transmit pilots for channel estimation at the HAP. The HAP estimates the uplink channels and utilizes the channel estimates to form the transmit beamformer for downlink transmission to all users, while sensors use the harvested energy to send data to the HAP simultaneously in the uplink. Further, we assume a massive antenna array at the HAP as a practical assumption~\cite{Kashyap:TWC:2016}. We obtain downlink-uplink sum-rate region by optimizing the energy beamformer and time-split parameter. Specifically, the downlink sum-rate is maximized by ensuring that the uplink sum-rate is above a certain threshold which is varied to maximum value the uplink sum-rate can take.

The main contributions of this paper are twofold.
\begin{itemize}
\item An optimum design, based on successive convex approximation (SCA) and semidefinite relaxation (SDR) for the beamformer and line search for the time-split parameter, is proposed.

\item We develop new tractable expressions for the achievable uplink and downlink rates in the large-antenna regime, along with approximating results that hold for any finite number of antennas for both perfect and imperfect channel state information (CSI) cases in the case of suboptimum maximum ratio transmission (MRT) beamforming. Our analysis reveals that for the perfect CSI case, in the limit of infinitely many receive antennas at the HAP, $\Ntx$, and energy harvesting, we can scale down the HAP transmit power proportionally to $1/\Ntx^2$.
\end{itemize}

\emph{Notation:} We use bold upper case letters to denote matrices, bold lower case letters to denote vectors. The superscripts  $(\cdot)^{T}$, and $(\cdot)^{\dag}$ stand for transpose and conjugate transpose respectively; the Euclidean norm of the vector, the Frobenius norm of the matrix, the trace, and the expectation are denoted by $\|\cdot\|$, $\|\cdot\|_F$, ${\rm tr} (\cdot)$, and ${\tt E}\left\{\cdot\right\}$ respectively; ${\rm vec}(\cdot)$ stands for the vectorization operation of the matrix; $\otimes$ denotes the matrix Kronecker product; $\qA \!=\!\diag\{\qA_1,\!\cdots\!,\!\qA_n\}$ stands for a block diagonal matrix and $\mathcal{CN}(\mu,\sigma^2)$ denotes a circular symmetric complex Gaussian random variable (RV) with mean $\mu$ and variance $\sigma^2$.
%***************************************************************************
%%***************************************************************************
\section{System Model}
%***************************************************************************
%%***************************************************************************
%as shown in Fig.~\ref{fig: system model}
\subsection{Energy Harvesting Network Topology}
We consider a FD HAP that simultaneously serves $\Kdl$  downlink users (cellular users) and $\Kul$ uplink users (sensor nodes) which are uniformly distributed in its coverage area. The HAP is equipped with a massive antenna array. The total number of antennas at the HAP is $N=\Nrx+\Ntx$ of which $\Nrx$ antennas are dedicated to the users and $\Ntx$ antennas are used for the sensors. The users and sensors are single antenna nodes while the sensors employ a rectanna each for energy harvesting. Each sensor node uses the harvested energy to power its subsequent uplink data transmission.
%%%%$$$$$$$$$$$$$$$$$$$$$$$$$$$$$$$$$$$$$$$$$$$$$$$$$$$$$$$$$$$$$$$$$$$$$$$$$$$$$$$$$$$$$$$$$$$$$$$$$$$$$$$$$$$$$$$$$$$$$$$$$
\vspace{-0.6em}
\subsection{Signal Transmission Model}
We consider frame-based transmissions over Rayleigh fading channels. The length of one frame is fixed to  $T$ seconds, which is assumed to be less than the coherence interval of the channel. Each frame is divided into two phases. In the first phase of time period $\alpha T$ ($0\leq \alpha \leq 1$), users transmit pilots for channel estimation at the HAP, while the HAP simultaneously transmits an energy signal to the sensor nodes. By channel reciprocity, the HAP obtains the downlink channels and then forms the beamformers for information transmission to the users during the interval $(1-\alpha)T$. Moreover, sensors transmit their data during the interval $(1-\alpha)T$ using the harvested energy.
\vspace{-0.6em}
\subsection{Uplink Channel Estimation for Users}
At the first phase of the $i$-th frame, all $\Kdl$ cellular users transmit pilot signal $p_k[i]$, $k\in\{1,\cdots.\Kdl\}$ with $\E{|p_k[i]|^2}=1$ to the HAP, while the HAP transmits energy signal $\qx_{\En,k}[i]$ to the $k$-th sensor node, given by

\vspace{-1.2em}
\small{
\begin{align}
\qx_{\En,k}[i] = \sqrt{P_\AP} \qw_{\En,k} s_\En[i],
\end{align}}\normalsize
where $\qw_{\En,k}\in \mathbb{C}^{\Ntx\times1}$ is the energy beamforming vector intended for $k$-th sensor with  $\|\qw_{\En,k}\|=1$. We assume that  $\E{|s_\En[i]|^2}=1$ so $P_\AP$ is the average transmit power of the HAP.  Denote the transmit pilot sequence of the users by $\qp[i] = [p_1[i],\cdots p_{\Kdl}[i]]^T$ and let $\qW_{\En} = [\qw_{\En,1}, \qw_{\En,2}, \cdots, \qw_{\En,\Kul}]$.
The received signal at the HAP is given by

\vspace{-1.2em}
\small{
\begin{align}\label{eq:phase1:y:HAP}
\qy_\AP [i] = \sqrt{P_{\dl}}\qG_{\AP\dl}\qp[i] + \sqrt{P_\AP}\qH_{\SI}\qW_{\En}\qs_{\En}[i] + \qn_\AP[i],
\end{align}}\normalsize
where $P_{\dl}$ denotes the transmit power of the users. $\qG_{\AP\dl}=[{\qg}_{\AP\dl,1},\cdots,{\qg}_{\AP\dl,\Kdl}]\in \mathbb{C}^{\Nrx\times \Kdl}$ is the channel matrix from the set of users to the HAP which is expressed as $\qG_{\AP\dl} = \qH_{\AP\dl}\qD_{\AP\dl}^{1/2}$, where the small-scale fading matrix $\qH_{\AP\dl}\in \mathbb{C}^{\Nrx\times \Kdl}$ has independent and identically distributed (i.i.d.) $\mathcal{CN}(0, 1)$ elements, while $\qD_{\AP\dl}$ is the large-scale path loss diagonal matrix whose $k$-th diagonal element is denoted by $\beta_{\AP\dl,k}$. The SI channel is represented by $\qH_{\SI}\in\mathbb{C}^{\Nrx\times\Ntx}$ with independent entries drawn from a $\mathcal{CN}(0,\Sap)$ distribution where $\Sap$ accounts for the residual SI power after SI suppression~\cite{Riihonen:JSP:2011,Himal:JSAC:2014}. $\qs_{\En}[i]\in\mathbb{C}^{\Kul\times1}$ is the energy symbol vector. The vector $\qn_\AP[i]\sim\mathcal{CN}(\textbf{0}, \Sn\qI_{\Nrx})$ accounts for additive white Gaussian noise (AWGN) at the HAP.

%\textbf{\emph{Channel estimation process: }}
Let $\tau$ number of pilot symbols.  During the training part, all users simultaneously transmit mutually
orthogonal pilot sequences, while HAP transmits the energy sequence of $\tau$ symbols. The pilot sequences used by the $\Kdl$ users can be represented by $\Plus \in\mathbb{C}^{\Kdl\times\tau}$ ($\tau\geq\Kdl$) which satisfies $\Plus \Plus^\dag = \qI_{\Kdl}$ . Let $\qS_{\En} \in\mathbb{C}^{\Kul\times\tau}$ be the energy sequence transmitted by the HAP. The received pilot signal at the HAP is given by

\vspace{-1.3em}
\small{
\begin{align}\label{eq:phase1:y:HAP:matrix}
\qY_\AP  &= \sqrt{P_{p}}\qG_{\AP\dl}\Plus + \sqrt{\tau P_\AP}\qH_{\SI}\qW_{\En}\qS_{\En} + \Nad\nonumber\\
& =  \Hbar\qX_p + \Nad,
\end{align}}\normalsize
where $P_p = \tau P_\dl$ is the transmit power of each pilot symbol; $\Hbar=\left[ {\qG}_{\AP\dl}, {\qH}_{\SI}\right]$ is the $\Nrx\times (\Kdl+\Ntx)$ channel matrix, $\qX_p=\left[ \sqrt{P_{p}}\Plus, \\ \sqrt{\tau P_\AP}\PhE \right]^T$ with $\PhE = \qW_{\En}\qS_{\En}$ is the $(\Kdl+\Ntx)\times \tau$ signal matrix, and $\Nad$ is the $\Nrx\times \tau$ matrix of HAP noise.

We assume that the HAP uses minimum mean-square-error (MMSE) estimation to estimate  the combined channel $\Hbar$. The linear MSE estimator of $\Hbar$ can be written as~\cite{Biguesh:TSP:2006}

\vspace{-1.3em}
\small{
\begin{align}\label{eq:MMSE est}
\Hbarhat=  \qY_\AP \left[\qX_p^\dag\qC_{\Hbar} \qX_p + \
\qI_{\tau}\right]^{-1}\qX_p^\dag\qC_{\Hbar},
\end{align}}\normalsize
where $\qC_{\Hbar}=\E{\Hbar^\dag \Hbar}=\diag\{\qD_{\AP\Ds}, \Sap\qI_{\Nrx}\}$. Moreover, the MMSE estimation error can be computed as $\text{MMSE}={\rm tr}\left( \left( \qC_{\Hbar} ^{-1}+ \frac{1}{\sigma_n^2}{\qX}_p{\qX}_p^\dag \right)^{-1}\right)$. It is known that the MMSE is minimized when ${\qX}_p{\qX}_p^\dag = \frac{\mathcal{P}}{\Kdl+\Ntx}\qI$ where $\mathcal{P} = \rm tr\{{\qX}_p{\qX}_p^\dag \}=\tau (P_\dl+ P_\AP)$~\cite{Biguesh:TSP:2006}. Therefore,

\vspace{-1.0em}
\small{
\begin{align}\label{eq:train:design}
%%\frac{1}{\tau} \Plus\Plus^\dag&={\qI}_{\Kdl},\nonumber\\
\frac{1}{\tau} \PhE \PhE^\dag={\qI}_{\Ntx},\quad
\PhE\Plus^\dag={\bf 0},\quad
\Plus\PhE^\dag={\bf 0}
\end{align}}\normalsize
The estimated channels can be decomposed by using MMSE properties as follows~\cite{Himal:JSAC:2014}

%======
\vspace{-1.2em}
\small{
\begin{align}\label{eq:MMSE decomp}
\Gad = \Gadhat + \Ead,\quad\HLI = \HLIhat + \ELI,
\end{align}}\normalsize
where $\Ead$ and $\ELI$ are the i.i.d Gaussian estimation error matrices of $\Gad$ and $\HLI$, respectively. From the property of MMSE channel estimation $\Gadhat$, $\Ead$, $\HLIhat$, and $\ELI$ are independent. Moreover,  the rows of $\Gadhat$, $\Ead$, $\HLIhat$, and $\ELI$
are mutually independent and distributed as $\mathcal{CN}(\bf{0}, \bf{\Omega}_{\AP\dl})$, $\mathcal{CN}(\bf{0}, \Dad-\bf{\Omega}_{\AP\dl})$, $\mathcal{CN}(\bf{0}, \bf{\Omega}_{\SI})$, and $\mathcal{CN}(\bf{0}, \Sap\qI - \bf{\Omega}_{\SI})$, respectively, where $\bf{\Omega}_{\AP\dl}$ and $\bf{\Omega}_{\SI}$    are diagonal matrices whose $k$-th diagonal elements are $\Sadk = \frac{\tau P_\dl\beta_{\AP\dl,k}^2}{1+ \tau P_\dl\beta_{\AP\dl,k}}$ and $\sigma^2_{\SI,k} = \frac{\tau P_\AP\sigma^4_{\SI}}{1+\tau P_\AP\Sap}$, respectively. The HAP-to-users channel CSI is obtained by using the reciprocity properties of the wireless channel as $\Gdahat= \Gadhat^{T}$
%%%%%%%%%%%%%%%%%%%%%%%%%%%%%%%%%%%%%%%%%%%%%%%%%%%%%%%%%%%%%
\vspace{-0.3em}
\subsection{Downlink Energy Transfer for Sensors}
%%%%%%%%%%%%%%%%%%%%%%%%%%%%%%%%%%%%%%%%%%%%%%%%%%%%%%%%%%%%%
During the first phase, the received signal at the $k$-th sensor can be expressed by

\vspace{-1.2em}
\small{
\begin{align}
 y_{\ul,k}[i] &=\sqrt{ P_\AP}\qg_{\AP\ul,k}^{T}\sum_{\ell=1}^{\Kul}\qw_{\En,\ell} s_\En[i]+\! \sqrt{P_\dl}\qg_{\dl\ul,k}^T \qp[i] +n_{\ul,k}[i],
\end{align}}\normalsize
%==============================
where $\qg_{\AP\ul,k} \in \mathbb{C}^{\Ntx\times1}$ and $\qg_{\dl\ul,k}\in \mathbb{C}^{\Kdl\times1}$ are the channel vectors from the HAP and the $\Kdl$ cellular users to the sensor node $k$, respectively. More precisely, $\qg_{\AP\ul,k}$ and $\qg_{\dl\ul,k}$ can be expressed as $\qg_{\AP\ul,k} = \sqrt{\beta_{\AP\ul,k}}\qh_{\AP\ul,k}$ and $ \qg_{\dl\ul,k}= \qD_{\dl\ul,k}^{1/2}\qh_{\dl\ul,k}$, respectively, where $\beta_{\AP\ul,k}$ denotes the large-scale path loss between HAP and sensor node $k$, the small-scale fading vectors $\qh_{\AP\ul,k}$ and $\qh_{\dl\ul,k}$ have i.i.d $\mathcal{CN}(0, 1)$ elements, while $\qD_{\dl\ul,k}$ is the large-scale path loss diagonal matrix whose $m$-th diagonal elements is denoted by ${\beta_{\dl\ul,k,m}}$ models the large-scale path loss between the $m$th user and $k$-th sensor, $n_{\ul,k}[i]\sim\mathcal{CN}(0,\Sn)$ denotes the AWGN at the $k$-th sensor.

Generally, the harvesting receiver will harvest energy from the whole signal $y_{\ul,k}[i] $. However, since the noise is negligible compared with the signal with large transmit power, and is thereby omitted in the harvested energy~\cite{RZhang:2013,Nasir.TWC.2013}. We further assume that the amount of energy harvested from the cellular users' transmissions is negligible due to their low transmit power.
Therefore, the $k$-th sensor node transmit power during the remaining $(1-\alpha)T$ time can be written as
%==========================================================

\vspace{-1.8em}
\small{
\begin{align}\label{eq:UL user power}
P_{\ul,k}&= \kappa  P_\AP\sum_{\ell=1}^{\Kul}|\qg_{\AP\ul,k}^{T}\qw_{\En,\ell}|^2
%,\quad k\in\{ 1,\cdots,\Kul\}
= \kappa  P_\AP\|\qg_{\AP\ul,k}^{T}\qW_{\En}\|^2,
%\quad k\in\{ 1,\cdots,\Kul\},
\end{align}}\normalsize
where $\kappa = \frac{\eta\alpha }{1-\alpha}$ and $0<\eta<1$ denotes the energy conversion efficiency.
%==========================================================
%==========================================================
\vspace{-1.1em}
\subsection{Uplink and Downlink Data Transmission}
%==========================================================
%==========================================================
The HAP uses the estimated channels to perform linear beamforming to transmit information to the users. At the same time, it receives data from the set of $\Kul$ sensors.

\emph{Uplink transmission:} At the second phase of the current time slot, the received signal at the HAP is separated into $\Kul$ streams by using the receive beamforming matrix $\qW_r = [\qw_{r,1},\cdots,\qw_{r,\Kul}]\in\mathbb{C}^{\Ntx\times\Kul}$ in which each column $\qw_{r,k}$ is the normalized receive beamforming vector assigned to the $k$-th sensor. Then, the received signal can be written as
%============================================================

\vspace{-1.2em}
\small{
\begin{align}\label{eq:UL data tx}
\qr_{\AP} [i] = \qW_r^{\dag}\qG_{\AP\ul}\qx_\ul[i] +
\sqrt{P_\AP}\qW_r^{\dag}{\qH}_{\SI}^{T}\qW_{t}\qx_{\dl}[i]+
\qW_r^{\dag}\qn_\AP[i],
\end{align}}\normalsize
%============================================================
where $\qG_{\AP\ul}=[{\qg}_{\AP\ul,1},\cdots,{\qg}_{\AP\ul,\Kul}]\in \mathbb{C}^{\Ntx\times \Kul}$ is the channel matrix from $\Kul$ sensor nodes to the HAP, $\qx_\ul[i] = [x_{\ul,1}[i],\cdots,x_{\ul,\Kul}[i]]^T$ with $\qQ_\ul \triangleq \E{\qx_\ul\qx_\ul^{\dag}} = \mathsf{diag}\{P_{\ul,1},\cdots,P_{\ul,\Kul}\}$ and $\qx_\dl[i] = [x_{\dl,1}[i],\cdots,x_{\dl,\Kdl}[i]]^T$  with $\E{\qx_\dl\qx_\dl^{\dag}} = \qI_{\Kdl}$ are the information-bearing signal of the sensor nodes and cellular users, respectively.  $\qW_t = [\qw_{t,1},\cdots,\qw_{t,\Kdl}]\in\mathbb{C}^{\Nrx\times\Kdl}$ denotes the downlink beamformer at the HAP in which each column $\qw_{t,k}$ is the normalized transmit beamforming vector assigned to the $k$-th user.
The $k$-th stream of $\qr_\AP [i]$ can be expressed as
%============================================================

\vspace{-1.3em}
\small{
\begin{align}\label{eq: UL data tx k-th}
r_{\AP,k} [i] &= \qw_{r,k}^{\dag}\qg_{\AP\ul,k} x_{\ul,k}[i]
+\sum_{\ell\neq k}^{\Kul}\qw_{r,k}^{\dag}\qg_{\AP\ul,\ell} x_{\ul,\ell}[i]+\nonumber\\
&\hspace{0.9em}
\sqrt{P_\AP}
\sum_{\ell=1}^{\Kdl}\qw_{r,k}^{\dag}{\qH}_{\SI}^T\qw_{t,\ell} x_{\dl,\ell}[i]
+
 \qw_{r,k}^{\dag}\qn_\AP[i].
\end{align}}\normalsize
%============================================================
From~\eqref{eq: UL data tx k-th}  the signal-to-interference plus noise ratio (SINR) corresponding to $k$-th sensor and observed at HAP is

\vspace{-1.3em}
\small{
\begin{align}\label{eq:SINR:HAP}
&\gamma_{\AP,k}(\qW_t,\qw_{r,k}, \qW_\En,\alpha) =\\
&\quad\frac{P_{\ul,k}|\qw_{r,k}^{\dag}\qg_{\AP\ul,k}|^2}
{\sum_{\ell\neq k}P_{\ul,\ell}|\qw_{r,k}^{\dag}\qg_{\AP\ul,\ell}|^2 +
P_\AP
\sum_{\ell=1}^{\Kdl}|\qw_{r,k}^{\dag}{\qH}_{\SI}^{T}\qw_{t,\ell}|^2+\Sn }.\nonumber
\end{align}}\normalsize

\emph{Downlink transmission:} The received signal from the HAP at the $\Kdl$ users is given by

\vspace{-1.3em}
\small{
\begin{align}\label{eq: DL data vec}
\qr_{\dl} [i] = \sqrt{P_\AP} {\qG}_{\AP\dl}^T\qW_{t}\qx_{\dl}[i]
                                                        + \qG_{\ul\dl}^{T} \qx_{\ul}[i]
                                             +\qn_{\dl}[i]
\end{align}}\normalsize
where $\qG_{\ul\dl}=[\qg_{\ul\dl,1},\cdots,\qg_{\ul\dl,\Kul}]\in\mathbb{C}^{\Kul\times\Kdl}$ represents channel matrix between the $\Kul$ sensor nodes and the $\Kdl$ users, i.e., $g_{\ul\dl,k,m} = [\qG_{\ul\dl}]_{km}$ is the channel coefficient
between the $m$-th sensor and the $k$-th user which can be written as $g_{\ul\dl,k,m} =\sqrt{\beta_{\ul\dl,k,m}} h_{\ul\dl,k,m}$ where $h_{\ul\dl,k,m}$ is the fast fading coefficient from the $m$-th sensor to the $k$-th user and $\sqrt{\beta_{\ul\dl,k,m}}$ models the large-scale path loss; $\qn_{\dl}[i]$ is the AWGN vector at $\Kdl$ users. The received signal at the $k$-th user can be written as

\vspace{-1.2em}
\small{
\begin{align}\label{eq: DL data k-th}
r_{\dl,k} [i] &= \sqrt{P_\AP} {\qg}_{\AP\dl,k}^T\qw_{t,k}x_{\dl,k}[i]\\
                                             &\hspace{0.5em}
                                             + \sqrt{P_\AP} {\qg}_{\AP\dl,k}^T \sum_{\ell\neq  k}^{\Kdl}\qw_{t,\ell}x_{\dl,\ell}[i]
                                             +
                                             \qg_{\ul\dl,k}^{T} \qx_{\ul}[i]
                                             + n_{\dl,k}[i],\nonumber
\end{align}}\normalsize
where $n_{\dl,k}[i]$ is the $k$-th element of $\qn_{\dl}[i]$. From~\eqref{eq: DL data k-th} the SINR at $k$-th user is

\vspace{-1em}
\small{
\begin{align}\label{eq:SINR:DL:user k}
&\gamma_{\mathsf{d},k}(\qW_t,\qW_\En,\alpha) =\\
&\qquad \frac{P_\AP\left|\qg_{\AP\dl,k}^T\qw_{t,k}\right|^2 }
{ P_\AP \sum_{\ell\neq  k}^{\Kdl}|\qg_{\AP\dl,k}^T \qw_{t,\ell}|^2 +
  \sum_{l=1}^{\Kul} P_{\ul,\ell} |g_{\ul\dl,k,\ell}|^2+
\Sn}.\nonumber
\end{align}}\normalsize
%==============================================
\vspace{-1em}
\section{Beamforming Optimization}
%==============================================
Our main purpose is to jointly design the receive, transmit and energy beamformers so that the downlink spectral efficiency is
maximized, while uplink spectral efficiency at the HAP is guaranteed to be above a certain value. This value is changed to maximum possible uplink spectral efficiency for obtaining the downlink-uplink sum-rate region.  As such, the
optimization problem can be expressed as

\vspace{-1.2em}
\small{
\begin{subequations}\label{eq:general opt}
\begin{align}
\max_{\qW_t,\qW_r, \qW_\En,~0\leq \alpha\leq 1} \hspace{0.7em}&  R_\mathsf{D}(\qW_t,\qW_\En,\alpha)\\
&\hspace{-5em}
\text{s.t} \hspace{4.2em}
R_\mathsf{U}(\qW_t,\qW_r, \qW_\En,\alpha)\!\geq\! \bar{R}_{\ul}\\
&\hspace{-7.5em}  \|\qW_t\|_F^2 =\Kdl,~\|\qW_r\|_F^2 =\Kul,~\|\qW_\En\|_F^2 =\Kul,
%&\hspace{-0.2em}  0\leq \alpha\leq 1,
\end{align}
\end{subequations}}\normalsize
where $R_\mathsf{D}=
(1-\alpha)\sum_{k=1}^{\Kdl}\log_2\left(1 + \gamma_{\mathsf{d},k}(\qW_t,\qW_\En,\alpha)\right)$ and $R_\mathsf{U}
= (1-\alpha)\!\sum_{k=1}^{\Kul}\!\log_2\left(1 + \gamma_{\AP,k}(\qW_t,\qw_{r,k},\qW_\En,\alpha)\right)$, and $\bar{R}_{\ul}$ is the minimum rate requirement for the sensor nodes. The optimization problem~\eqref{eq:general opt} is a complicated non-convex optimization problem with respect to (w.r.t) the beamforming vectors and $\alpha$. By considering good performance and low complexity of maximum ratio combination (MRC) and MRT in multiuser MIMO systems, we investigate the optimum energy beamforming matrix $\qW_\En$ and time-split parameter $\alpha$. We note that these precoders are asymptotically optimal in massive MIMO~\cite{Debbah.JSAC.2013}. Substituting $\qw_{r,k}^\MRC\!\!=\!\!\frac{\qg_{\AP\ul,k}}{\| \qg_{\AP\ul,k}\|}$ and $\qw_{t,k}^\MRT\!\!=\!\!\frac{\qg_{\AP\dl,k}^*}{\| \qg_{\AP\dl,k}\|}$ into~\eqref{eq:SINR:HAP} and~\eqref{eq:SINR:DL:user k}, and using the law of large numbers~\cite{Ngo:TWC:.2013}, when $\Nrx$ grows large,~\eqref{eq:general opt} can be approximated as
%==========================================================

\vspace{-1.3em}
\small{
%\begin{subequations}
\begin{align}\label{optimum P1}
\hspace{-0em}\max_{\qW_\En, 0\leq \alpha\leq1}\hspace{-.3em}& \!(1\!-\!\alpha)
\!\!\sum_{k=1}^{\Kdl}\!
\log_2\!\!
\left(\!1\!\!+\!\frac{P_\AP \|\qg_{\AP\dl,k}\|^2}
{\frac{\eta P_\AP\alpha}{(1-\alpha)}{\rm tr}\left( {\tilde \qH}_{\AP\ul, k}\qW_{\En}\qW_{\En}^\dag {\tilde \qH}_{\AP\ul, k}^\dag\right)\!+\!\Sn}  \!\right)\!,\nonumber\\
&\hspace{-2.8em}
\text{s.t}
\hspace{0.8em}
\sum_{k=1}^{\Kul}\!
\log_2\!\left(1\!+\!
\frac{c_k\alpha}{(1\!-\alpha)} \qh_{\AP\ul, k}^T \qW_{\En} \qW_{\En}^\dag  \qh_{\AP\ul, k}^{*} \right)
\!\geq\!\! \frac{\bar{R}_{\ul}}{(1\!-\!\alpha)},\nonumber\\
 &\hspace{-0.8em}
 \|\qW_\En\|_F^2 = \Kul,
\end{align}}\normalsize
%\end{subequations}
%==========================================================
where \small{${\tilde \qH}_{\AP\ul, k}\!\!=\!\!\left[\!\! \sqrt{\beta_{\AP\ul,1}} | g_{\ul\dl,k,1}| \qh_{\AP\ul, 1}^T;
 \!\cdots\!;\sqrt{\beta_{\AP\ul,\Kul}} | g_{\ul\dl,k,\Kul}| \qh_{\AP\ul, \Kul}^T \!\right]$ }\normalsize  and $c_k = \frac{\kappa P_\AP}{(P_\AP \Sap +\Sn)} \beta_{\AP\ul,k}\|\qg_{\AP\ul,k}\|^2$.
%==========================================================
%==========================================================
While problem~\eqref{optimum P1} is non-convex and its global optimum solutions cannot be found in polynomial time, it can be efficiently solved as shown in the following. Introducing an auxiliary variable $\tau_k$,~\eqref{optimum P1} is expressed as
%==========================================================

\vspace{-1.2em}
\small{
%\begin{subequations}
\begin{align}\label{optimum P1.1}
\max_{\qW_\En,\tau_k, 0\leq \alpha\leq1}\hspace{-0em}& (1-\alpha)
\sum_{k=1}^{\Kdl}
\log_2\
\left(1+P_\AP \|\qg_{\AP\dl,k}\|^2 \tau_k\right),\nonumber\\
&\hspace{-2.9em}
\text{s.t}
\hspace{1.5em}
\tau_k\leq
\frac{1}
{\frac{\eta P_\AP\alpha}{1-\alpha}{\rm tr}\!\left(\! {\tilde \qH}_{\AP\ul, k}\qW_{\En}\qW_{\En}^\dag {\tilde \qH}_{\AP\ul, k}^\dag\!\right)\!+\!\Sn}, \forall k,
\nonumber\\
&\hspace{-5em}
\sum_{k=1}^{\Kul}
\log_2\!\left(1+\!
\frac{c_k\alpha}{(1-\alpha)} \qh_{\AP\ul, k}^T \qW_{\En} \qW_{\En}^\dag  \qh_{\AP\ul, k}^{*} \right)
\!\geq\! \frac{\bar{R}_{\ul}}{(1\!-\alpha)},\nonumber\\
&\hspace{-5em}
\|\qW_\En\|_F^2 = \Kul.
\end{align}
%\end{subequations}
}\normalsize
%=====================================
Note that $\qh_{\AP\ul, k}^T \qW_{\En} \qW_{\En}^\dag  \qh_{\AP\ul, k}^{*}$ can be rewritten as
%==========================================================

\vspace{-1.2em}
\small{
\begin{align}\label{eq:hWWh}
\hspace{0em}{\rm tr}\left(\! \qh_{\AP\ul, k}^{*}\qh_{\AP\ul, k}^T \qW_{\En} \qW_{\En}^\dag  \!\right)
&\stackrel{a}{=}\!\left( \!\left(\!\qh_{\AP\ul, k}^{*}\qh_{\AP\ul, k}^T\otimes\qI\right)\!{\rm vec}\left(\qW_{\En}\right)\right)^\dag\!\!{\rm vec}\left(\qW_{\En}\right)\nonumber\\
&\stackrel{c}=\bar{\qw}_{\En}^\dag
\left( \qh_{\AP\ul, k}\qh_{\AP\ul, k}^\dag\otimes\qI\right)
\bar{\qw}_{\En},
\end{align}}\normalsize
%==============================================
where (a) follows by using the matrix identity ${\rm tr}(\qA\qB^\dag) = ({\rm vec}(\qB))^\dag{\rm vec}(\qA)$ and ${\rm vec}(\qA\qX\qB) = (\qA^{T}\otimes\qB){\rm vec}(\qX)$, respectively and ${\rm vec}\left(\qW_{\En}\right) \triangleq\bar{\qw}_{\En}$ in (c). Similarly, we get
%==========================================================

\vspace{-1.0em}
\small{
\begin{align}\label{eq:HWWH}
\hspace{-0.5em} {\rm tr}\left(\! {\tilde \qH}_{\AP\ul, k}\qW_{\En}\qW_{\En}^\dag {\tilde \qH}_{\AP\ul, k}^\dag\!\right) \!=\!\bar{\qw}_{\En}^\dag
\left( {\tilde \qH}_{\AP\ul, k}^T{\tilde \qH}_{\AP\ul, k}^*\otimes\qI\right)
\bar{\qw}_{\En}.
\end{align}}\normalsize

We now apply the SDR technique by using a positive-semidefinite matrix $\bar{\qW}_\En = \bar{\qw}_\En \bar{\qw}_\En^\dag$ and relaxing the rank-constraint on $\bar{\qW}_\En$. By using~\eqref{eq:hWWh} and~\eqref{eq:HWWH}, the optimization problem~\eqref{optimum P1.1} can be re-formulated as
%=====================================

\vspace{-1.25em}
\small{
%\begin{subequations}
\begin{align}\label{optimum P1.2}
\hspace{-1em}\max_{\bar{\qW}_\En,\tau_k,0\leq\alpha\leq1}\hspace{0.2em}& (1-\alpha)
\sum_{k=1}^{\Kdl}
\log_2\!
\left(1+P_\AP \|\qg_{\AP\dl,k}\|^2 \tau_k\right),\nonumber\\
&\hspace{-3em}
\text{s.t}
\hspace{2em}
\frac{\eta P_\AP\alpha}{(1-\alpha)}
{\rm tr}\left(\bar{\qW}_{\En}\qA_k\!\right)
\!+\!\Sn\leqslant\frac{1}{\tau_k}, \forall k,
\nonumber\\
&\hspace{-2em}
\sum_{k=1}^{\Kul}
\log_2\!\left(1\!+\!
\frac{c_k\alpha}{(1-\alpha)}
{\rm tr}\left(\bar{\qW}_{\En}
\qB_k\!\right)\! \right)
\!\geq\! \frac{\bar{R}_{\ul}}{(1\!-\alpha)},\nonumber\\
&\hspace{-2em}
{\rm tr}(\bar{\qW}_\En) = \Kul, \quad\bar{\qW}_{\En}\succeq 0,
\end{align}
%\end{subequations}
}\normalsize
where $\bar{\qA}_k = \left( {\tilde \qH}_{\AP\ul, k}^T{\tilde \qH}_{\AP\ul, k}^*\otimes\qI\right)$ and  $\bar{\qB}_k =\left( \qh_{\AP\ul, k}\qh_{\AP\ul, k}^\dag\otimes\qI\right)$. The optimization problem~\eqref{optimum P1.2} is still non-convex (even w.r.t $\{\bar{\qW}_\En,\tau_k\}$) due to the second set of constraint in which there is a term $(\tau_k)^{-1}$. In the following, we show that the problem~\eqref{optimum P1.2} can be solved efficiently by finding optimum $\bar{\qW}_\En$ for a given $\alpha$. Since $\alpha$ is scalar valued, its optimum solution can be ascertained by using one-dimensional search w.r.t. $\alpha$. Let ${\bar{\tau}_k} = (\tau_k)^{-1}$ and $f_k(\bar{\tau}_k) = \log_2(\bar{\tau}_k)$. Since $f_k(\bar{\tau}_k)$ is concave, we have
%==========================================================

\vspace{-1.1em}
\small{
\begin{align}\label{eq:Ttylor}
f_k(\bar{\tau}_k)\leq f_k(\bar{\tau}_{k,0}) + \frac{\partial f_k(\bar{\tau}_{k,0})}{\partial \bar{\tau}_{k} } (\bar{\tau}_{k} - \bar{\tau}_{k,0}).
\end{align}}\normalsize
%To this end, the optimization problem becomes
To this end, using~\eqref{eq:Ttylor}, the objective function in~\eqref{optimum P1.2} can be approximated by its lower bound. Consequently, for a given $\alpha$,~\eqref{optimum P1.2} becomes
%=====================================

\vspace{-1.3em}
\small{
%\begin{subequations}
\begin{align}\label{optimum P1.4}
\max_{\bar{\qW}_\En,\bar{\tau}_k}\hspace{1em}&
\sum_{k=1}^{\Kdl}
\log_2\
\left({\bar{\tau}_k}+P_\AP \|\qg_{\AP\dl,k}\|^2\right) - \nonumber\\
&\hspace{1.7em}\frac{1}{\log(2)}
\sum_{k=1}^{\Kdl}\left(\log({\bar{\tau}_{k,0}}) + \frac{1}{{\bar{\tau}_{k,0}}}({\bar{\tau}_{k}}-{\bar{\tau}_{k,0}})\right),\nonumber\\
&\hspace{-2.5em}
\text{s.t}
\hspace{1.5em}
\frac{\eta P_\AP\alpha}{(1-\alpha)}
{\rm tr}\left(\bar{\qW}_{\En}\bar{\qA}_k\right)
+\!\Sn\leqslant {\bar{\tau}_k},\quad\forall k,
\nonumber\\
&\sum_{k=1}^{\Kul}
\log_2\!\left(1+\!
\frac{c_k\alpha}{(1-\alpha)}
{\rm tr}\left(\bar{\qW}_{\En}\bar{\qB}_k
 \right)\right)
\!\geq\! \frac{\bar{R}_{\ul}}{(1\!-\alpha)},\nonumber\\
 &{\rm tr}(\bar{\qW}_\En) = \Kul, \quad\bar{\qW}_{\En}\succeq 0.
\end{align}
%\end{subequations}
}\normalsize
%==========================================================
The optimization problem~\eqref{optimum P1.4} successively approximates~\eqref{optimum P1.2} as an SDR, for a given $\alpha$. The obtained rank-one $\bar{\qW}_\En$ (or its approximated rank-one solution~\cite{Batu:TSP:2009}) is then used to recover $\qW_\En$. The proposed optimization method is outlined in Algorithm 1.
%%%%%%%%%%%%%%%%%%%%%%%%%%%%%%%%%%%%%%%%%%%%%%%
%\vspace{-0.5em}
 \begin{algorithm}
 \caption{ The proposed optimization scheme}
 \begin{algorithmic}
  \renewcommand{\algorithmicrequire}{\textbf{Step 1:} }
 \REQUIRE Initialize $\alpha$:
 Choose $\alpha$ from its grid: $\alpha \in [0, 1)$.
 \renewcommand{\algorithmicrequire}{\textbf{Step 2:} }
 \REQUIRE For given $\alpha$:
      \WHILE{not converged}
        \STATE Start with initial $[\bar{\tau}_{1,0},\cdots,\bar{\tau}_{\Kdl,0}]$.
        \STATE Obtain the transmit beamformer $\bar{\qW}_\En$ using~\eqref{optimum P1.4},

   \STATE Update $[\bar{\tau}_{1,0},\cdots,\bar{\tau}_{\Kdl,0}]$ with the solutions of~\eqref{optimum P1.4}.

  \ENDWHILE\label{euclidendwhile}
   \IF { $\bar{\qW}_\En$ is rank-one}
     \STATE $\bullet$ Take $\bar{\qw}_\En$ as the  eigenvector corresponding to maximum eigenvalue of  $\bar{\qW}_\En$ and scale with $\sqrt{\Kul}$.
     \STATE $\bullet$ Recover ${\qW}_\En$ from $\bar{\qw}_\En$.
 \ELSE
    \STATE Approximate rank-one solution~\cite{Batu:TSP:2009}.
 \ENDIF
  \STATE Take another value of $\alpha$ and go to \textbf{Step 2}.
  \renewcommand{\algorithmicrequire}{\textbf{Step 3:} }
 \REQUIRE Take $\alpha$ and ${\qW}_\En$ that maximize objective function.
 \end{algorithmic}
 \end{algorithm}
%%%%%%%%%%%%%%%%%%%%%%%%%%%%%%%%%%%%%%%%%%%%%%%%%%%%%%%%%%%%%%
%==================================
\vspace{-0.5em}
\section{Achievable Rate Analysis}
%==================================
We carry out  the achievable rate analysis with perfect and imperfect CSI in this section. We focus on the case in which MRC/MRT processing is considered for uplink/downlink information transfer and MRT beamformer is employed for energy transfer as motivated in~\cite{Long:TWC:2016}.
%=============================================
%=============================================
%~\cite{Hassibi:IT:.2003}
\vspace{-1em}
\subsection{Uplink Transmission}
%=============================================
%=============================================
\emph{1) Perfect CSI:} By invoking~\eqref{eq:SINR:HAP} and by using a standard bound based on the worst-case uncorrelated additive noise for the  perfect CSI case,  the achievable uplink rate of the $k$-th sensor can be expressed as
~\eqref{eq: Rulk PCSI1} at the top of the page,
\small{
\bigformulatop{22}{ \vskip-0.5cm
\begin{align}\label{eq: Rulk PCSI1}
\Ruk&= (1-\alpha){\tt E}\left\{\log_2\left(1+
\frac{\frac{\kappa P_\AP}{\Ntx^2}
\left(\sum_{j=1,j\neq k}^{\Kul}
\frac{|\qg_{\AP\ul,k}^{T} \qg_{\AP\ul,j}^{*}|^2}{\|\qg_{\AP\ul,j}\|^2}  \|\qg_{\AP\ul,k}\|^2 + \|\qg_{\AP\ul,k}\|^4\right)
}
{\frac{\kappa P_\AP}{\Ntx^2}
\sum_{\ell\neq k}^{\Kul}
\left(\sum_{j=1, j\neq \ell}^{\Kul} \frac{|\qg_{\AP\ul,\ell}^{T} \qg_{\AP\ul,j}^{*}|^2}{\|\qg_{\AP\ul,j}\|^2} |\hat{g}_\ell|^2 +
\|\qg_{\AP\ul,\ell}\|^2|\hat{g}_\ell|^2 \right)
+
\frac{ P_\AP}{\Ntx^2}
\sum_{\ell=1}^{\Kdl}
|\tilde{g}_\ell|^2
+
\frac{\Sn}{\Ntx^2} }\right)\right\}
\end{align}
}}\normalsize
where \small{$\tilde{g}_\ell =\frac{\qg_{\AP\ul,k}^\dag\HLI^{T}\qg_{\AP\dl,\ell}^*}{\|\qg_{\AP\ul,k}\| \| \qg_{\AP\dl,\ell}\|}$ }\normalsize and \small{$\hat{g}_\ell =\frac{\qg_{\AP\ul,k}^{\dag}\qg_{\AP\ul,\ell}}{\|\qg_{\AP\ul,k}\|}$. }\normalsize When $\Ntx$ grows large, \small{$\frac{1}{\Ntx^2}|\qg_{\AP\ul,k}^{T} \qg_{\AP\ul,j}^{*}|^2\rightarrow 0$ }\normalsize for \small{$k\neq j$ }\normalsize and \small{$\frac{1}{\Ntx}\|\qg_{\AP\ul,\ell}\|^2\!\!\rightarrow\!\!\beta_{\AP\ul,\ell}$. }\normalsize Hence,~\eqref{eq: Rulk PCSI1} can be approximated as
%============================

\small{
\setcounter{equation}{23}
\begin{align}\label{eq: Rulk PCSI3}
\hspace{2em}\Ruk&\approx (1-\alpha){\tt E}\left\{\log_2\left(1+\right.\right.\\
&\left.\left.\frac{\kappa P_\AP \Ntx
\beta_{\AP\ul,k}\|\qg_{\AP\ul,k}\|^2
}
{\kappa P_\AP \Ntx
\sum_{\ell\neq k}^{\Kul}
\beta_{\AP\ul,\ell}|\hat{g}_\ell|^2
+
 P_\AP
\sum_{\ell=1}^{\Kdl}
|\tilde{g}_\ell|^2
+
\Sn }\right)\right\}.\nonumber
\end{align}}\normalsize
%========================================
\bigformulatop{30}{ \vskip-0.5cm
\small{
\begin{align}\label{eq: Rdlk PCSI1}
\Rdk&= \!(1-\!\alpha){\tt E}
\left\{\log_2\left(\!1\!+
\frac{ \frac{P_\AP}{\Ntx^2} \|\qg_{\AP\dl,k}\|^2
}
{ \frac{P_\AP}{\Ntx^2}
\sum_{\ell\neq k}^{\Kdl}
|\breve{g}_\ell|^2
\!+
\kappa \frac{P_\AP}{\Ntx^2}
\sum_{\ell=1}^{\Kul}
\left( \sum_{j=1, j\neq \ell}^{\Kul}
\frac{| \qg_{\AP\ul,\ell}^T \qg_{\AP\ul,j}^*|^2}{ \| \qg_{\AP\ul,j}\|^2} + \| \qg_{\AP\ul,\ell}\|^2 \right)
|g_{\ul\dl,k,\ell}|^2
\!+
 \frac{\Sn}{\Ntx^2}}\right)\right\},
\end{align}
}}\normalsize
%========================================
\begin{proposition}\label{UL:perfect}
With perfect CSI and MRC/MRT processing at the HAP, the uplink achievable rate from the $k$-th sensor can be approximated as

\vspace{-1.2em}
\small{
\begin{align}\label{eq: Rulk PCSI APX1}
\Ruk&\!\approx\!(1-\alpha) \!\!\int_{0}^{\infty}\!\!
\left(\frac{1}{1 + P_\AP \Sap z}\right)^{\Kdl}\!\!\!
\left(1 \!-\! \left(\frac{1}{1 \!+\! \varphi_k z}\!\right)^{\Ntx}\!\right)\!\nonumber\\
&\hspace{0em}\times\prod_{\ell=1, \ell\neq k}^{\Kul}\!\left(\!\frac{1}{1+\varphi_\ell z}\!\right)
\frac{e^{-\Sn z}}{z}
dz,
\end{align}}\normalsize
where  $\varphi_\ell=\kappa P_\AP \Ntx\beta_{\AP\ul,\ell}^2$ and $\varphi_k=\kappa P_\AP \Ntx\beta_{\AP\ul,k}^2$.
\end{proposition}
%\begin{proof}
\emph{proof:} The proof is omitted due to space limitations.
%\end{proof}

To the best of the authors' knowledge, the integral in~\eqref{eq: Rulk PCSI APX1} does not admit a closed-form
expression. However this integral can be efficiently evaluated numerically. Alternatively, we can use the following closed-form lower bound.
%for uplink achievable rate.
%=======================================================
\begin{proposition}\label{UL:perfect:lower bound}
Assume that the AP has perfect CSI, and $\Ntx\geq 2$, the uplink achievable rate from the $k$-th sensor for the MRC/MRT processing scheme at the HAP can be lower bounded as:

\vspace{-1.2em}
\small{
\begin{align}\label{eq:lower:UL:PCSI}
\Ruk&\geq  \Rukhat= (1-\alpha)\times \\
& \log_2\left(1+\frac{\kappa P_\AP
\beta_{\AP\ul,k}^2
(\Ntx+2)(\Ntx-1)}
{\kappa P_\AP
\Ntx
\sum_{\ell\neq k}^{\Kul}
\beta_{\AP\ul,\ell}^2
+\Kdl P_\AP\sigma_\SI^2
+\Sn}\right).\nonumber
\end{align}}\normalsize
Moreover, if $P_\AP=\frac{E_\AP}{\Ntx^2}$ and $\Ntx$ becomes infinity, then

\vspace{-1.1em}
\small{
\begin{align}\label{eq:SNRul with MRT}
\Rukhat \rightarrow (1-\alpha)\log_2\left(1+\kappa\beta_{\AP\ul,k}^2 \frac{E_\AP}{\Sn } \right),~\Ntx \rightarrow \infty.
\end{align}}\normalsize
\end{proposition}
%=======================================================
%\begin{proof}
\emph{proof:} The proof follows from the convexity of \small{$\log_2\left(1 + \frac{1}{x}\right)$ }\normalsize
and using Jensen's inequality.
%\end{proof}

\begin{Remark}
Proposition~\ref{UL:perfect:lower bound} indicates that with perfect CSI at the HAP and a
large $\Ntx$, the uplink performance of the system with transmit power per user of $P_\AP = \frac{E_\AP}{\Ntx^2 }$ is equal to the performance of a single-input single-output system with transmit power $E_\AP$ without any fading. By using a large number of HAP antennas and energy harvesting, we can scale down the transmit power proportionally to $1/\Ntx^2.$
\end{Remark}

\emph{2) Imperfect CSI:}  In this case, imperfect CSI of the SI and user-to-HAP channels is available at the HAP, whereas the HAP has perfect CSI of the sensor-to-HAP channels. Therefore, the received signal associated with the $k$-th sensor can be expressed as
%============================================================

\vspace{-1.4em}
\small{
\begin{align}\label{eq: UL data imp k-th}
r_{\AP,k} [i] &= \qw_{r,k}^{\dag}\qg_{\AP\ul,k} x_{\ul,k}[i]
+\sum_{\ell\neq k}^{\Kul}\qw_{r,k}^{\dag}\qg_{\AP\ul,\ell} x_{\ul,\ell}[i]\\
&\hspace{0.2em}
+\sqrt{P_\AP}
\sum_{\ell=1}^{\Kdl}\qw_{r,k}^{\dag}(\hat{\qH}_{\SI}^T -\ELI^{T})\hat{\qw}_{t,\ell} x_{\dl,\ell}[i]
+
 {\qw}_{r,k}^{\dag}\qn_\AP[i].\nonumber
\end{align}}\normalsize
%============================================================
Since the HAP knows its own transmit signal $\hat{\qw}_{t}x_{\dl}[i]$ and the MMSE estimation of the SI channel, the SI term \small{$\sqrt{P_\AP}\qw_{r,k}^{\dag}\hat{\qH}_{\SI}^T \hat{\qw}_{t,\ell} x_{\dl,\ell}[i]$ }\normalsize in~\eqref{eq: UL data imp k-th} can be canceled~\cite{Riihonen:JSP:2011}. Accordingly, the achievable uplink rate of the $k$-th sensor node with the imperfect CSI is given~\eqref{eq: Rulk PCSI1} by replacing $\tilde{g}_\ell$ with \small{ $\acute{g}_\ell=\frac{\qg_{\AP\ul,k}^\dag\ELI^{T}\hat{\qg}_{\AP\dl,\ell}^*}{\|\hat{\qg}_{\AP\ul,k}\| \| \hat{\qg}_{\AP\dl,\ell}\|}$. }\normalsize

%============================================
\begin{proposition}\label{UL:IPCSI:Asymp}
Assume that the HAP has imperfect CSI of the user-to-HAP channels and perfect CSI of the HAP-to-sensor channels, the uplink achievable rate from the $k$-th sensor for MRC/MRT processing scheme at the HAP can be lower bounded as

\vspace{-1.4em}
\small{
\begin{align}\label{eq:lower:UL:ECSI}
\RukhatIP&\geq  \Rukhat= (1-\alpha)\times \\
& \log_2\left(1+\frac{\kappa P_\AP
\beta_{\AP\ul,k}^2
(\Ntx+2)(\Ntx-1)}
{\kappa P_\AP
\Ntx
\sum_{\ell\neq k}^{\Kul}
\beta_{\AP\ul,\ell}^2
+\frac{\Kdl P_\AP\sigma_\SI^2}{\tau P_\AP \Sap + 1}
+\Sn}\right).\nonumber
\end{align}}\normalsize
Moreover, if $P_\AP=\frac{E_\AP}{\Ntx}$ and $\Ntx$ grows without bound, then

\vspace{-1.4em}
\small{
\begin{align}\label{eq:SNRul with MRT}
\RukhatIP \rightarrow (1-\alpha)\log_2\left(1+ \frac{E_\AP^2}{\Sn } \kappa\beta_{\AP\ul,k}^2\right),~\Ntx \rightarrow \infty.
\end{align}}\normalsize
\end{proposition}

%\begin{proof}
\emph{proof:} The proof is omitted due to space limitations.
% and thus has been omitted.
%\end{proof}
%=============================================
%=============================================
\vspace{-1.5em}
\subsection{Downlink Transmission}
%=============================================
%=============================================
\emph{2) Perfect CSI:} We derive the achievable downlink rate of the genie receiver, i.e., $k$-th user knows $\qg_{\AP\dl,k}^T\qw_{t,\ell}$, $\ell=1,\cdots,\Kdl$. In this case, the ergodic achievable downlink rate of the $k$-th user can be written as~\eqref{eq: Rdlk PCSI1} at the top of the page where  $\breve{g}_\ell = \frac{ \qg_{\AP\dl,k}^T \qg_{\AP\dl,\ell}^*}{ \| \qg_{\AP\dl,\ell}\|}$. By the law of large numbers, when $\Ntx$ grows large, we obtain that $\frac{1}{\Ntx^2}| \qg_{\AP\ul,\ell}^T \qg_{\AP\ul,j}^*|^2\rightarrow 0$, for $\ell\neq j$ and $\frac{1}{\Nrx}\| \qg_{\AP\ul,\ell}\|^2\rightarrow \beta_{\AP\ul,\ell}$.  Therefore, the achievable downlink rate of the $k$-th user can be written as
%==============================

\vspace{-1.4em}
\small{
\setcounter{equation}{31}
\begin{align}\label{eq: Rdlk PCSI2}
\Rdk&\approx (1-\alpha){\tt E}
\left\{\log_2\left(1+\right.\right.\\
&\!\!\left.\left.\frac{ P_\AP \|\qg_{\AP\dl,k}\|^2
}
{
P_\AP
\sum_{\ell\neq k}^{\Kdl}
|\breve{g}_\ell|^2
\!+\!
\kappa P_\AP\Ntx
\sum_{\ell=1}^{\Kul}
 \beta_{\AP\ul,\ell}
|g_{\ul\dl,k,\ell}|^2
\!+
\Sn}\right)\!\right\}\!.\nonumber
\end{align}}\normalsize
%==================
Conditioned on $\qg_{\AP\dl,k}$, $\breve{g}_\ell$ is Gaussian RV
with zero mean and unit variance which does not depend on $\qg_{\AP\dl,k}$. Therefore, ˜$\breve{g}_\ell$ is Gaussian distributed and independent of $\qg_{\AP\dl,k}$, $\breve{g}_\ell\sim \mathcal{CN}(0, \beta_{\AP\dl,k})$.
Accordingly,  following proposition states  the uplink achievable rate.

\begin{proposition}\label{DL:perfect:Asymp}
The achievable downlink rate of the $k$-th user with perfect CSI can be approximated as

\vspace{-1.1em}
\small{
\begin{align}\label{eq: Rulk PCSI APX}
\Ruk&\!\approx\!(1-\alpha) \int_{0}^{\infty}\!\!
\!\left(\frac{1}{1 + \psi_k z}\right)^{\!\!\Kdl-1}
\!\!\!\left(1 \!-\left(\!\frac{1}{1 \!+\! \psi_k z}\right)^{\Nrx}
\right)\nonumber\\
&\hspace{0em}\times\prod_{\ell=1, \ell\neq k}^{\Kul}\!\left(\!\frac{1}{1+\psi_\ell z}\!\right)
\frac{e^{-\Sn z}}{z}
dz,
\end{align}}\normalsize
where  $\psi_k=P_\AP \beta_{\AP\dl,k}^2$ and $\psi_\ell=\kappa P_\AP \Ntx\beta_{\AP\ul,\ell}\beta_{\ul\dl,k,\ell}$.
\end{proposition}
%\begin{proof}
\emph{proof:} The proof is omitted due to space limitations.
%\end{proof}
\begin{proposition}\label{DL:perfect:lower bound}
Assume that the HAP has perfect CSI, and $\Ntx\geq 2$, the downlink achievable rate of the $k$-th user with MRC/MRT processing scheme at the HAP can be lower bounded as

\vspace{-1.1em}
\small{
\begin{align}\label{eq:lower:DL:PCSI}
\Rdk&\geq  \Rdkhat= (1-\alpha)\log_2\left(1+ \right. \\
&\left.\frac{ P_\AP
\beta_{\AP\dl,k}
(\Nrx-1)}
{(\Kdl-1) P_\AP \beta_{\AP\dl,k} + \kappa P_\AP
\Ntx
\sum_{\ell=1}^{\Kul}
\beta_{\AP\ul,\ell}
\beta_{\ul\dl,k,\ell}
+\Sn}\right).\nonumber
\end{align}}\normalsize
\end{proposition}
%=======================================================
%\begin{proof}
\emph{proof:} The proof follows from the convexity of $\log_2\left(1 + \frac{1}{x}\right)$
and using Jensen's inequality.
%\end{proof}

%====================================================
%====================================================
\emph{2) Imperfect CSI:} With the above communication scheme, the HAP has channel estimates while the users do not
have any channel estimate. To this end, we provide an ergodic achievable rate based on the techniques developed in~\cite{Marzetta:TWC:.2011}. Therefore, we decompose the received $r_{\dl,k} [i]$ as

\vspace{-1.1em}
\small{
\begin{align}\label{eq: DL data k-th decom}
r_{\dl,k} [i] &= \sqrt{P_\AP} {\tt E}\left\{ {\qg}_{\AP\dl,k}^T\qw_{t,k} \right\}x_{\dl,k}[i]+ \tilde{n}_{\dl,k}[i].
\end{align}}\normalsize
%==================================================
In~\eqref{eq: DL data k-th decom} the effective noise is defined as
%==================================================

\vspace{-1.1em}
\small{
\begin{align}\label{eq: DL data k-th effective noise}
\tilde{n}_{\dl,k}[i] &\!=\!
 \sqrt{P_\AP} \left({\qg}_{\AP\dl,k}^T\qw_{t,k}\!-\!{\tt E}\left\{ {\qg}_{\AP\dl,k}^T\qw_{t,k} \right\}\right)x_{\dl,k}[i]
                                           + \\
                                           &\hspace{0.8em}
                                           \sqrt{P_\AP} {\qg}_{\AP\dl,k}^T \sum_{\ell\neq  k}^{\Kdl}\qw_{t,\ell}x_{\dl,\ell}[i]
+ \qg_{\ul\dl,k}^{T} \qx_{\ul}[i]
                                             + n_{\dl,k}[i].\nonumber
\end{align}}\normalsize
%==================================================
The average effective channel ${\tt E}\left\{ {\qg}_{\AP\dl,k}^T\qw_{t,k} \right\}$ can be perfectly learned at the users. Thus, the expectation is known as it only depends on the channel distribution and is not related to the instantaneous channel. However, the additive noise $\tilde{n}_{\dl,k}[i]$ is neither independent nor Gaussian. We use the fact
%result in~\cite{Hassibi:IT:.2003}
that shows that worst-case uncorrelated additive noise is independent Gaussian noise with the same variance to derive the achievable downlink rate as

\vspace{-1em}
\small{
\begin{align}\label{eqn:Rdk_impCSI}
\RdkIP  &\!=\!
(1-\alpha)\log_2\left(1\!+\! \frac{P_\AP\left| {\tt E}\left\{\qg_{\AP\dl,k}^T\qw_{t,k}\right\}\right|^2 }
{ P_\AP\text{Var}(\qg_{\AP\dl,k}^T\qw_{t,k})+ \mathsf{I}_k+
\Sn}\right).
\end{align}
}\normalsize
where \small{$\mathsf{I}_k =P_\AP \sum_{\ell\neq  k}^{\Kdl}{\tt E}\left\{|\qg_{\AP\dl,k}^T \qw_{t,\ell}|^2\right\} + \sum_{l=1}^{\Kul} {\tt E}\left\{P_{\ul,\ell}| g_{\ul\dl,k,\ell}|^2\right\}$. }\normalsize
%%==================================================

%====================================================
\begin{proposition}\label{Propos:Rd:IP}
Using the imperfect CSI from MMSE estimation, the achievable downlink rate of the $k$-th user is given by
%===============================================

\vspace{-1.1em}
\small{
\setcounter{equation}{37}
\begin{align}\label{eq:Rdk ImpCSI}
&\RdkIP =\log_2\left(1+\right.\\
&\left.\frac{ P_\AP \Nrx^2\sigma_{\AP\dl,k}^4}
{ P_\AP\Nrx\Sadk\sum_{\ell=1}^{\Kdl} \beta_{\AP\dl,\ell}
 \!+\!\kappa P_\AP\Ntx \Kul \!\sum_{\ell=1}^{\Kul}\!\beta_{\AP\ul,\ell}\beta_{\ul\dl,k,\ell}\! +\! \Sn }\!\right).\nonumber
\end{align}}\normalsize
%===============================================
%where $\Sadk = \frac{\tau P_\dl\beta_{\AP\dl,k}^2}{1+ \tau P_\dl\beta_{\AP\dl,k}}$.
\end{proposition}
%\begin{proof}
\emph{proof:} The proof is omitted due to space limitations.
%See Appendix~\ref{Appendix:Proof RDIP}.
%\end{proof}
%===================================
\vspace{-0.5em}
\section{Numerical and Simulation Results}
%===================================
We now evaluate the performance the FD HAP system using simulations where the accuracy of the presented analytical results are also verified. Unless mentioned otherwise, we have set $\eta=0.5$, $\Kul=3$, $\Kdl=5$, $\Sn=1$ and $\Sap=1$.
``PCSI" and ``ECSI" represent the results with perfect CSI and estimated CSI, respectively.

\begin{figure}[t]
\centering
\vspace{0em}
\includegraphics[width=85mm, height=52mm]{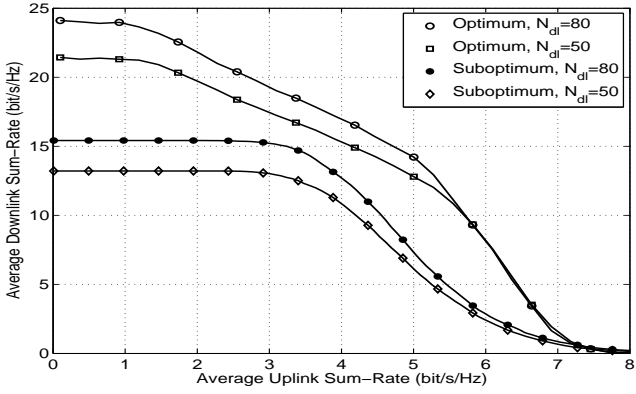}
\vspace{-1.8em}
\caption{Uplink-downlink sum-rate region for the proposed Algorithm 1 and suboptimum schem with $P_\AP=20$ dB and $\Ntx=10$.}
\vspace{-1em}
\label{fig1:rate:region}
\end{figure}

Fig. \ref{fig1:rate:region} shows the rate regions obtained with the proposed Algorithm 1 and suboptimal scheme.
As expected, the proposed Algorithm 1 performs better than the suboptimum scheme. This observation can be explained as follows. The MRT energy beamformers used in the suboptimal scheme try to maximize power transfer to each sensor without considering that sensors can transmit  their uplink data in the following slot with high powers if they harvest more energy in the energy harvesting phase.  As such, sensor nodes can produce significant interference to downlink data transmission from HAP.  On the other hand, the proposed scheme tries to maintain a tradeoff between energy transfer to sensor nodes and the interference they generate, due to the harvested energy,  to downlink transmission from HAP.
 %%%%$$$$$$$$$$$$$$$$$$$$$$$$$$$$$$$$$$$$$$$$$$$$$$$$$$$$$$$$$$$$$$$$$$$$$$$$$$$$$$$$$$$$$$$$$$$$$$$$$$$$$$$$$$$$$$$$$$$$$$$$$
%%%%$$$$$$$$$$$$$$$$$$$$$$$$$$$$$$$$$$$$$$$$$$$$$$$$$$$$$$$$$$$$$$$$$$$$$$$$$$$$$$$$$$$$$$$$$$$$$$$$$$$$$$$$$$$$$$$$$$$$$$$$$

\begin{figure}[t]
\centering
\vspace{-1.1em}
\includegraphics[width=85mm, height=57mm]{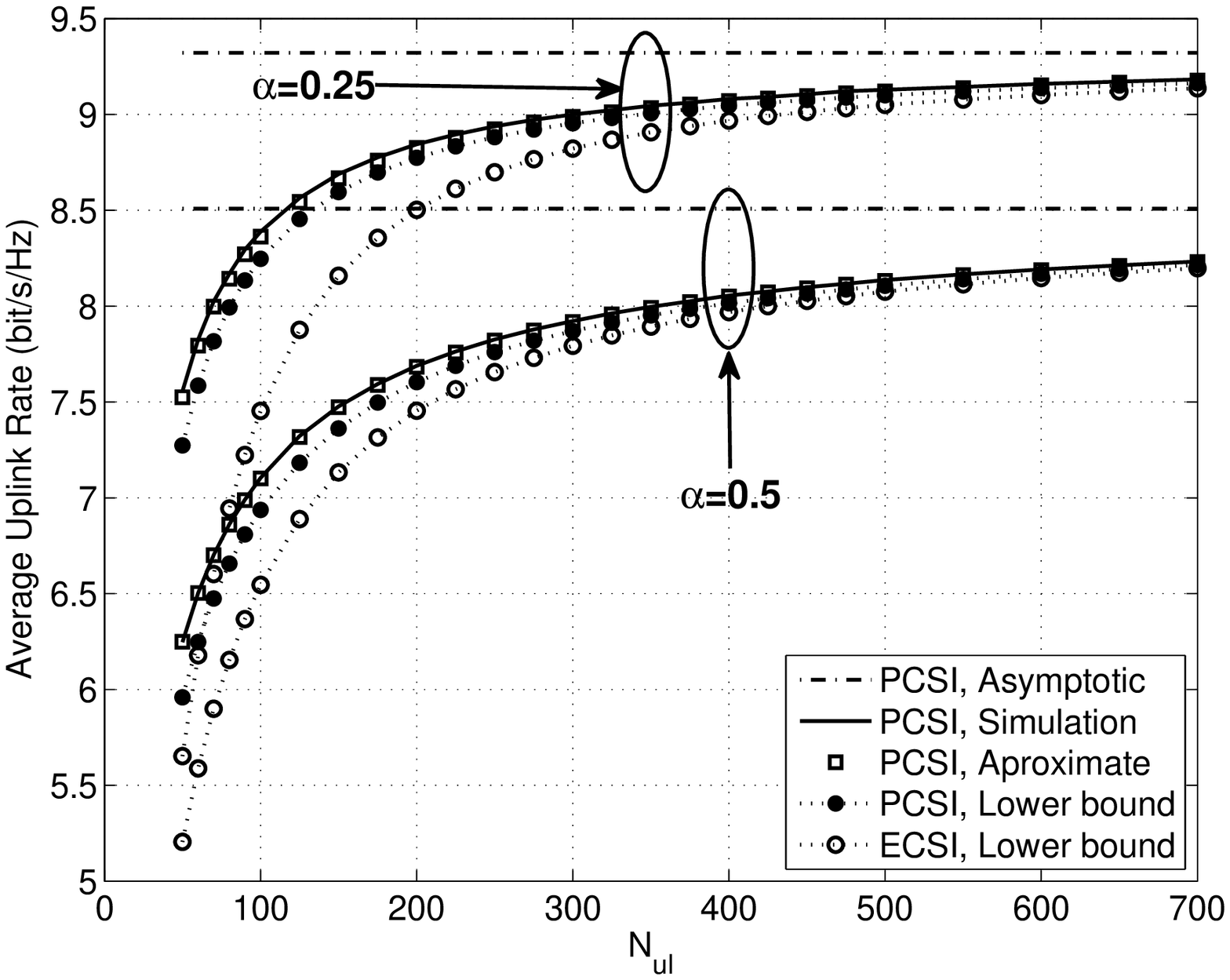}
\vspace{-2.3em}
\caption{Acheivable uplink sum-rate versus $\Ntx$ with $\Nrx=50$.}
\vspace{-1.5em}
\label{fig: ULrate_Ant}
\end{figure}

In Fig.~\ref{fig: ULrate_Ant} the uplink sum-rate, defined as $C_\ul =  \sum_{k=1}^{\Kul} \Ruk$, is plotted versus $\Ntx$ with two different values of $\alpha$. When the number of antennas is infinite, the asymptotic results with PCSI is given by~\eqref{eq:SNRul with MRT}. The lower bounds on the information rate with PCSI and ECSI are calculated by~\eqref{eq:lower:UL:PCSI} and~\eqref{eq:lower:UL:ECSI}, respectively. It is observed that the simulation results for both PCSI and ECSI tend to the asymptotic result when the number of antennas increases. However, when $\alpha$ is increased the gap between the asymptotic result and simulation result increases since longer energy harvesting time increases the harvested energy and consequently intra-sensor interference.
%===================================
\section{Conclusion}
%===================================
We have analyzed the performance of a massive MIMO-enabled uplink/downlink information and energy transfer system with the time-switching protocol. Assuming perfect CSI, we have maximized the downlink sum-rate of a set of downlink users through the joint energy beamformer and time-split factor design at the HAP, by ensuring that the uplink sum-rate of a set of energy-limited sensors is above a certain threshold. In addition, for both perfect and imperfect CSI scenarios, we derived expressions for the achievable uplink and downlink rate when the number of HAP antennas grows without bound. We further provided asymptotic results that hold for any finite number of antennas. Accordingly, we established a scaling law showing that by using a large number of HAP antennas and energy harvesting, the HAP transmit power can scale down as $1/\Ntx^2$ and $1/\Ntx$ for perfect and imperfect CSI, respectively.
 %%%%$$$$$$$$$$$$$$$$$$$$$$$$$$$$$$$$$$$$$$$$$$$$$$$$$$$$$$$$$$$$$$$$$$$$$$$$$$$$$$$$$$$$$$$$$$$$$$$$$$$$$$$$$$$$$$$$$$$$$$$$$
%%%%$$$$$$$$$$$$$$$$$$$$$$$$$$$$$$$$$$$$$$$$$$$$$$$$$$$$$$$$$$$$$$$$$$$$$$$$$$$$$$$$$$$$$$$$$$$$$$$$$$$$$$$$$$$$$$$$$$$$$$$$$
%=======================================
%\bibliographystyle{IEEEtran}

%%=======================================
%\vspace{-1em}
%\bibliographystyle{IEEEtran}
%\bibliography{IEEEabrv,Ref_MassiveMIMO}

\begin{thebibliography}{1}

\bibitem{Ashutosh:JSAC:2014}
A. Sabharwal \emph{et al.}, ``In-band full-duplex wireless: Challenges and opportunities'' \emph{
IEEE J. Sel. Areas Commun.}, vol. 32, pp. 1637-1652, Sep. 2014.

\bibitem{Riihonen:JSP:2011}
T. Riihonen, S. Werner, and R. Wichman, ``Mitigation of loopback self-interference in full-duplex MIMO relays,'' \emph{ IEEE Trans. Signal Process.}, vol. 59, pp. 5983-5993, Dec. 2011.

\bibitem{Sabharwal:TWC2012}
M. Duarte, C. Dick, and A. Sabharwal, ``Experiment-driven characterization
of full-duplex wireless systems,'' \emph{IEEE Trans. Wireless
Commun.}, vol. 11, pp. 4296-4307, Dec. 2012.



\bibitem{R.Zhang:JSAC:.2015}
G. Yang, C. K. Ho, R. Zhang, and Y. L. Guan, ``Throughput optimization
for massive MIMO systems powered by wireless energy transfer,"
\emph{IEEE J. Sel. Areas Commun.}, vol. 33, pp. 1640-1650, Aug. 2015.

\bibitem{Long:TWC:2016}
L. Zhao, X. Wang, and K. Zheng, ``Downlink hybrid information and
energy transfer with massive MIMO," \emph{IEEE Trans. Wireless Commun.},
vol. 15, pp. 1309-1322, Feb. 2016.

%
\bibitem{HJu:TCOM:2014}
H. Ju and R. Zhang, ``Optimal resource allocation in full-duplex
wireless-powered communication network," \emph{IEEE Trans. Commun.},
vol. 62, pp. 3528-3540, Oct. 2014.

%
\bibitem{Yongbo:2016}
Y. Cheng, P. Fu, Y. Chang, B. Li, and X. Yuan, ``Joint power and time
allocation in full-duplex wireless powered communication networks,"
\emph{Mobile Information Systems,} 2016.

%
\bibitem{Nguyen:Eusipco2016}
V.-D. Nguyen, H. V. Nguyen, G.-M. Kang, H. M. Kim, and O.S.
Shin, ``Sum rate maximization for full duplex wireless-powered
communication networks," in \emph{Proc. European Signal Process. Conf.
(EUSIPCO’16)}, Budapest, Hungary, Aug./Sep. 2016, pp. 798-802.
%
%
\bibitem{Kashyap:TWC:2016}
S. Kashyap, E. Bj\"{o}rnson, and E. G. Larsson, ``On the feasibility of
wireless energy transfer using massive antenna arrays," \emph{IEEE Trans.
Wireless Commun.}, vol. 15, pp. 3466-3480, May 2016.
%
\bibitem{Himal:JSAC:2014}
H. Q. Ngo, H. A. Suraweera, M. Matthaiou, and E. G. Larsson, ``Multipair
full-duplex relaying with massive arrays and linear processing,"
\emph{IEEE J. Sel. Areas Commun.}, vol. 32, pp. 1721-1737, June 2014.
%
\bibitem{Biguesh:TSP:2006}
M. Biguesh and A. B. Gershman, ``Training-based MIMO channel
estimation: a study of estimator tradeoffs and optimal training signals,"
\emph{IEEE Trans. Signal Process.}, vol. 54, pp. 884-893, Mar. 2006.

\bibitem{RZhang:2013}
R. Zhang and C. K. Ho, ``MIMO broadcasting for simultaneous wireless
information and power transfer," \emph{IEEE Trans. Wireless Commun.} ,
vol. 12, pp. 1989-2001, May 2013.

\bibitem{Nasir.TWC.2013}
A. A. Nasir, X. Zhou, S. Durrani, and R. A. Kennedy, ``Relaying
protocols for wireless energy harvesting and information processing,"
\emph{IEEE Trans. Wireless Commun.}, vol. 12, pp. 3622-3636, July 2013.


\bibitem{Debbah.JSAC.2013}
J. Hoydis, S. ten Brink, and M. Debbah, ``Massive MIMO in the
UL/DL of cellular networks: How many antennas do we need?" \emph{IEEE
J. Sel. Areas Commun.}, vol. 31, no. 2, pp. 160-171, Feb. 2013.

\bibitem{Ngo:TWC:.2013}
H. Q. Ngo, E. G. Larsson, and T. L. Marzetta, ``Energy and spectral
efﬁciency of very large multiuser MIMO systems," \emph{IEEE Trans.
Commun.}, vol. 61, pp. 1436-1449, Apr. 2013.

\bibitem{Batu:TSP:2009}
B. K. Chalise and L. Vandendorpe, ``MIMO relay design for
multipoint-to-multipoint communications with imperfect channel state
information," \emph{IEEE Trans. Signal Process.}, vol. 57, pp. 2785-2796,
July 2009.


\bibitem{Marzetta:TWC:.2011}
J. Jose, A. E. Ashikhmin, T. L. Marzetta, and S. Vishwanath, ``Pilot
contamination and precoding in multi-cell TDD systems," \emph{IEEE Trans.
Wireless Commun.}, vol. 10, pp. 2640-2651, Aug. 2011.


\end{thebibliography}
%%=======================================
%=======================================
\end{document}